\documentstyle[trans]{rs}
\series{A} \year{1996}
\begin{document}

\title[Continuous elastic phase transitions]
      {Continuous elastic phase transitions in pure and disordered crystals}

\author[F. Schwabl and U.C. T\"auber]
       {Franz Schwabl $^1$ and Uwe Claus T\"auber $^{1,2}$}

\affiliation{$^1$ Institut f\"ur Theoretische Physik,
                  Physik--Department der TU M\"unchen,\\
                  James--Franck--Stra\ss e, D--85747 Garching, Germany \\
             $^2$ Lyman Laboratory of Physics, Harvard University,\\
                  Cambridge, Massachusetts 02138, U.S.A.
                  \thanks{Present address:
                  Department of Physics -- Theoretical Physics,
                  1 Keble Road, University of Oxford, Oxford OX1 3NP, U.K.}}
\maketitle

\centerline{To appear in: {\bf Phil. Trans. R. Soc. Lond. A (1996)}}
\bigskip \bigskip

\begin{abstract}

We review the theory of second--order (ferro--)elastic phase transitions, where
the order parameter consists of a certain linear combination of strain tensor
components, and the accompanying soft mode is an acoustic phonon.
In three--dimensional crystals, the softening can occur in one-- or
two--dimensional soft sectors. The ensuing anisotropy reduces the effect of
fluctuations, rendering the critical behaviour of these systems classical for a
one--dimensional soft sector, and classical with logarithmic corrections in
case of a two--dimensional soft sector.
The dynamical critical exponent is $z = 2$, and as a consequence the sound
velocity vanishes as $c_s \propto | T - T_c |^{1/2}$, while the phonon damping
coefficient is essentially temperature--independent.
Even if the elastic phase transition is driven by the softening of an optical
mode linearly coupled to a transverse acoustic phonon, the critical exponents
retain their mean--field values.
Disorder may lead to a variety of precursor effects and modified critical
behaviour. Defects that locally soften the crystal may induce the phenomenon of
local order parameter condensation.
When the correlation length of the pure system exceeds the average defect
separation $n_{\rm D}^{-1/3}$, a disorder--induced phase transition to a state
with non--zero average order parameter can occur at a temperature
$T_c(n_{\rm D})$ well above the transition temperature $T_c^0$ of the pure
crystal. Near $T_c^0$, the order--parameter curve, susceptibility, and
specific heat appear rounded. For $T < T_c(n_{\rm D})$ the spatial
inhomogeneity induces a static central peak with finite $q$ width in the
scattering cross section, accompanied by a dynamical component that is confined
to the very vicinity of the disorder--induced phase transition.
\end{abstract}

\newpage


\section{Introduction}

\noindent
Generally, displacive structural phase transformations (for reviews, see Bruce
\& Cowley 1981 and Rao \& Rao 1987) can be divided into two different groups,
namely distortive and elastic phase transitions. At distortive phase
transitions some of the ions or molecular groups in the crystal's elementary
cell are displaced with respect to each other. The corresponding displacement
field serves as an appropriate order parameter for the transition, which in the
case of a second--order transition is thus characterized by the softening of an
optical phonon mode that becomes overdamped in the vicinity of the transition.
Ionic displacements may lead to the appearance of a macroscopic polarization,
i.e., to ferro-- or antiferroelectric behaviour. Famous examples of this group
of structural phase transitions are the perovskites with the high--temperature
simple--cubic ${\rm ABO_3}$ crystal structure; e.g.: the ferroelectric
${\rm BaTiO_3}$, the antiferroelectric ${\rm NaNbO_3}$, and ${\rm SrTiO_3}$,
which displays an antiferrodistortive transition accompanied by the softening
of a zone--boundary phonon (see L\"uthi \& Rehwald 1981). At these distortive
transformations the elastic degrees of freedom constitute merely secondary
variables, their interaction with the critical order parameter fluctuations
being quadratic in the order parameter and linear in the strains.

At an elastic phase transition (also called ferroelastic), on the other hand,
the crystallographic unit cell undergoes an elastic deformation, as is depicted
in Fig.~\ref{elcons} for ${\rm NaOH}$ (for a review, see Wadhawan 1982). Thus
the order parameter in this case is a certain linear combination of appropriate
components of the strain tensor. In the majority of cases the crystal undergoes
a shear deformation, and the accompanying soft mode is the corresponding
transverse acoustic phonon. ${\rm KH_2PO_4}$ ${\rm (KDP)}$, ${\rm KCN}$, and
${\rm Fe_3O_4}$, for example, display a first--order elastic phase
transformation, while in K-Na-tartrate, ${\rm Nb_3Sn}$, ${\rm TeO_2}$,
${\rm KH_3(SeO_3)_2}$, ${\rm LaP_5O_{14}}$, and ${\rm DyVO_4}$ the transition
is continuous (second--order), which implies that (at least) one of the sound
velocities vanishes at the transition (L\"uthi \& Rehwald 1981; Cummins 1982).
As a consequence of the crystal anisotropy, the sound velocities depend on the
direction of propagation, and thus the soft--phonon velocity does not vanish
throughout the entire Brillouin zone, but only in one or several so--called
soft sectors which may be one-- or two--dimensional. E.g., for the martensitic
transformation in the A-15 compounds ${\rm Nb_3Sn}$ and ${\rm V_3Si}$ the
transverse acoustic phonons propagating along the face diagonals of the
Brillouin zone with sound velocity $c_s = \sqrt{(c_{11}-c_{12}) / 2 \rho}$
soften, see Fig.~\ref{sndvel} (Rehwald {\em et al.} 1972); in ${\rm KCN}$ the
elastic constant $c_{44}$ vanishes at the transition, and the soft subspaces
are two--dimensional (Hauss\"uhl 1973; Knorr {\em et al.} 1985). We remark that
the actual situation may be quite complex; e.g., in the ferroelectric
${\rm KDP}$ the z component of the electric dipole moment couples linearly to
the $\varepsilon_{12}$ shear deformation (Brody \& Cummins 1974). Consequently,
although caused by the ordering of hydrogen--bonding protons, the transition
becomes ultimately an elastic phase transformation with vanishing shear modulus
$c_{66}$. Similar cases, where the instability itself occurs for an optical
mode that then drives the elastic transition, are ${\rm LaP_5O_{14}}$
(Fox {\em et al.} 1976) and h-${\rm BaTiO_3}$ (Yamaguchi {\em et al.} 1995),
amongst others. Yet there are also cases of ``pure'' elastic phase transitions,
where local fluctuations render the crystal unstable, e.g., in ${\rm NaOH}$
(Bleif {\em et al.} 1971). The other possibility, realized in the so--called
isostructural elastic phase transformations, is related to those elastic
stability limits where all transverse and necessarily all longitudinal sound
velocities remain finite. For instance, if the bulk modulus of an isotropic or
cubic elastic medium vanishes, only the macroscopic uniform dilatation and
gradient modes soften, but none of the phonons (Khmel'nitskii 1975; this kind
of behaviour was observed in ${\rm Ce}$, see Poniatovskii 1958).

\medskip
\centerline{\fbox{\parbox[b]{5.5cm}{Figure \ref{elcons} and Figure \ref{sndvel}
                                    near here}}}
\medskip

In the present brief overview, we shall focus on the theory of continuous
elastic phase transitions accompanied by the softening of an acoustic phonon.
We shall rely on the Ginzburg--Landau free--energy functional that was
established by Cowley (1976) and Folk {\em et al.} (1976 {\em a,b}), who also
derived the Langevin equations of motion for the soft acoustic phonons
(Folk {\em et al.} 1979). We are going to investigate the statics and
soft--phonon dynamics of second--order elastic phase transitions in pure
systems in \sect2 and \sect3, respectively; for earlier reviews, see
(Schwabl 1980) and (Schwabl 1985). As a consequence of the anisotropy of the
elastic system, fluctuations are reduced and the (upper) critical dimension
$d_c(m) = 2 + m/2$ (for the case of a $m$--dimensional soft sector). This is
smaller than $d_c = 4$ for a $\phi^4$ model, describing either an Ising or
Heisenberg magnet, or distortive structural transitions with short--range
interactions. If the theory is phrased in terms of deformations, the
Hamiltonian is seen to be equivalent to a spin system with long--range uniaxial
dipolar interactions, which again explains why the corresponding critical
dimension is lower (see \sect2). The case where the elastic transition is
actually driven by the linear coupling of transverse acoustic modes to a
softening optical phonon will be briefly discussed in \sect4.

The at the time surprising observation of extremely narrow central peaks in the
scattering cross--section, near both distortive (Riste {\em et al.} 1971;
Shapiro {\em et al.} 1972) and elastic transitions (Shirane \& Axe 1971) has
prompted various theoretical studies investigating the influence of lattice
defects on the statics and dynamics of structural phase transformations (e.g.,
Folk \& Schwabl 1974; Halperin \& Varma 1976; H\"ock \& Thomas 1977;
Schmidt \& Schwabl 1977, 1978; H\"ock {\em et al.} 1979;
Sasv\'ari \& Schwabl 1982; Weyrich \& Siems 1984; 
Wiesen {\em et al.} 1987, 1988). For reviews of the experimental facts, see
(M\"uller 1979) and (Fleury \& Lyons 1982); some key theoretical results are
collected in (Bruce \& Cowley 1981). Recently, these studies, which are
typically concerned with distortive transformations and single--defect
properties or conclusions derived by linear superposition thereof, were
extended to elastic systems (Schwabl \& T\"auber 1991 {\em a}), and the methods
developed for distortive crystals with a truly finite concentration of
impurities (Schwabl \& T\"auber 1991 {\em b}) were applied to the highly
anisotropic elastic structural phase transitions (Bulenda {\em et al.} 1995).
These developments and their consequences for a possible explanation of the 
central--peak phenomenon are the subject of \sect5. In the concluding \sect6 we
summarize by contrasting elastic phase transitions with distortive structural 
transformations, and also discussing some aspects of the related first--order
martensitic transformations.


\section{Static properties in pure systems}

\subsection{Ginzburg--Landau free energy functional}

In order to study elastic phase transitions, we expand the elastic free energy
in terms of the strain tensor $\varepsilon_{ik}$ (Einstein's sum convention is
employed here and in the following),
\begin{eqnarray}
{\cal F}[\{ \varepsilon_{ik} \}] &&= \int \biggl[
        {1 \over 2} c^{(2)}_{iklm} \, \varepsilon_{ik} \varepsilon_{lm} +
        {1 \over 2} d^{(2)}_{iklmrs}
                \left( {\partial \over \partial x_{r}} \varepsilon_{ik} \right)
                \left( {\partial \over \partial x_{s}} \varepsilon_{lm} \right)
                                                        \label{elasfe} \\
        &&\qquad + {1 \over 3!} c^{(3)}_{iklmrs} \,
                \varepsilon_{ik} \varepsilon_{lm} \varepsilon_{rs} +
                {1 \over 4!} c^{(4)}_{iklmrsuv} \, \varepsilon_{ik}
                \varepsilon_{lm} \varepsilon_{rs} \varepsilon_{uv} + \ldots
                                        \biggr] \, d^3x \quad , \nonumber
\end{eqnarray}
where the strain tensor is defined via the derivatives of the displacement
field $u_i$,
\begin{equation}
\varepsilon_{ik} = {1 \over 2} \left(
        {\partial u_i \over \partial x_k} + {\partial u_k \over \partial x_i} +
        {\partial u_l \over \partial x_i} {\partial u_l \over \partial x_k}
                                                \right) \quad , \label{strain}
\end{equation}
and the displacement field $u_i({\bf x})$ itself can be expanded in terms of
the Fourier--space normal coordinates $Q_{{\bf k},\lambda}$
($\lambda = 1,2,3$),
\begin{equation}
u_i({\bf x}) = {1 \over \sqrt{N M}} \sum_{{\bf k},\lambda} e^{i{\bf k}{\bf x}}
        e_i({\bf k},\lambda) \, Q_{{\bf k},\lambda} \quad . \label{displc}
\end{equation}
Here, ${\bf e}({\bf k},\lambda)$ denotes the polarization vector, $M$ the mass
and $N$ the number of the unit cells.

The coefficients $c^{(m)}$ in Eq.~(\ref{elasfe}) are the isothermal elastic
constants of order $m$. In addition to the terms contained in the standard
expansion of the free energy in elasticity theory, we explicitly take into
account contributions involving gradients of the strain tensor. Contributions
of this form were introduced in so--called generalized elasticity theory, and
can be derived from a lattice--dynamical model by taking the long--wavelength
limit and retaining terms up to the fourth order in the wave vector $k$ (see
Krumhansl 1968; Mindlin 1968; Kunin 1968). These gradient terms are of course
essential for the mathematical description of spatial fluctuations and
short--range inhomogeneities.

\medskip
\centerline{\fbox{\parbox[b]{5.5cm}{Figure \ref{orttet} and Figure \ref{cubhex}
                                    near here}}}
\medskip

The point--group symmetry now determines the number of independent elastic
constants $c^{(2)}_{iklm}$ and, via the stability limits, the possible elastic
phase transformations (see Aubry \& Pick 1971, Liakos \& Saunders 1982). E.g.,
in the orthorhombic system one finds elastic phase transitions with vanishing
$c_{44}$, $c_{55}$, and $c_{66}$ (we apply the Voigt notation). The
corresponding wave vectors and polarization vectors are shown in
Fig.~\ref{orttet}{\em a}. In case of higher symmetries softening may occur in
one-- or two--dimensional sectors (if $c_{44} \rightarrow 0$) of Fourier space.
The possible acoustic soft modes in tetragonal, cubic, and hexagonal crystals
are illustrated in Figs.~\ref{orttet} and \ref{cubhex}, respectively. Since
hexagonal crystals are elastically isotropic with respect to the six--fold c
axis, the transverse phonons with wave vector and polarization perpendicular to
the c axis are degenerate and soften with $c_{66} = (c_{11}-c_{12})/2$ going to
zero.

If there are third--order invariants which lead to a cubic term in the soft
phonon mode, the transition will be of first order. The third--order invariants
allowed by symmetry were tabulated by Brugger (1965). E.g., in cubic systems
with $c_{44} \rightarrow 0$ there is a third--order term of the form
$\varepsilon_{12} \varepsilon_{23} \varepsilon_{13}$; third--order terms are
also present for $c_{11}-c_{12} \rightarrow 0$ in cubic and hexagonal systems.
The theory described in the following is applicable, if third--order terms are
either not present at all, or if they are sufficiently small in order that the
ensuing first--order character of the transition will only be noticeable in the
immediate vicinity of the transition temperature $T_c$, and the phase
transformation can effectively be considered as continuous. The
high--temperature phases and vanishing combination of elastic constants for the
possible elastic phase transitions are listed in table~\ref{elptra}, along with
the corresponding strain components
[$e_2 = (\varepsilon_{11}-\varepsilon_{22})/\sqrt{2}$,
$e_3 = (\varepsilon_{11} + \varepsilon_{22} - 2 \varepsilon_{33})/\sqrt{6}$],
dimensionality $m$ of the soft sectors, and the third--order invariants. For a
listing of physical examples for these transformations, see (Folk {\em et al.}
1979; L\"uthi \& Rehwald 1981; Cummins 1982).

\medskip
\centerline{\fbox{\parbox[b]{3cm}{Table \ref{elptra} near here}}}
\medskip

A few remarks are in place here concerning the soft--mode spectrum and the
characteristic Hamiltonian or free--energy density following from the expansion
(\ref{elasfe}), as to be investigated in the following \sect2$\,b\,$ and
\sect3.

(i) The sound velocity vanishes only in one-- or two--dimensional subspaces. In
the vicinity of these directions the sound velocity is non--zero but small, and
it is important for the theory to include all phonons with wavevectors
${\bf k}$ in a finite sector around the $m$--dimensional soft subspace.

(ii) The gradient terms stemming from generalized elasticity theory prevent the
sound frequency from vanishing throughout the entire Brillouin zone, and
replace the linear by a quadratic dispersion precisely at the critical
temperature $T_c$.

(iii) For the study of critical phenomena, we shall disregard non--critical
modes and only retain the normal coordinate of the soft acoustic phonon.

(iv) We shall discard odd anharmonic terms, thus restricting the applicability
of the theory to situations where these are prohibited by symmetry; this
applies to orthorhombic and tetragonal crystals, or to cases where the cubic
terms are sufficiently small such that the phase transition can be regarded as
of nearly second order.

\subsection{Critical statics}

We are now ready to address the static critical behaviour near elastic phase
transitions. In the framework of renormalization group theory, one considers a
$d$--dimensional crystal with an $m$--dimensional soft subspace (in real
systems, of course, $d = 3$ and $m = 1$ or $m = 2$). Accordingly, we decompose
the $d$--dimensional wave vector ${\bf k}$ into its $m$--dimensional ``soft''
components ${\bf q}$ and the $(d-m)$--dimensional ``stiff'' components
${\bf p}$; ${\bf k} = ({\bf q},{\bf p})$. The Hamiltonian following from
Eqs.~(\ref{elasfe}), (\ref{displc}) and the discussion at the end of the
preceding Section then reads in Fourier space (Folk {\em et al.} 1976
{\em a,b})
\begin{eqnarray}
H[\{ Q_{\bf k} \}] &&=
        \frac12 \int d^dk \left( r p^2 + p^4 + q^2 \right) | Q_{\bf k} |^2 
                                                                \nonumber \\
        &&\qquad + u \int d^dk_1 \ldots d^dk_4 \, v({\bf k}_1,\ldots,{\bf k_4})
                \, Q_{{\bf k}_1} Q_{{\bf k}_2} Q_{{\bf k}_3} Q_{{\bf k}_4}
                                                        \quad , \label{hamilt}
\end{eqnarray}
where $r \propto T - T_c$ is assumed (in the following, we shall omit the
polarization indices $\lambda$). We will investigate the following two models:
\begin{eqnarray}
{\rm I}:  &&v({\bf k}_1,{\bf k}_2,{\bf k}_3,{\bf k_4}) = p_1 p_2 p_3 p_4 \,
      \delta({\bf k}_1+{\bf k}_2+{\bf k}_3+{\bf k_4}) \quad , \label{model1} \\
{\rm II}: &&v({\bf k}_1,{\bf k}_2,{\bf k}_3,{\bf k_4}) =
                ({\bf p}_1 {\bf p}_2) ({\bf p}_3 {\bf p}_4) \,
      \delta({\bf k}_1+{\bf k}_2+{\bf k}_3+{\bf k_4}) \quad . \label{model2}
\end{eqnarray}
The characteristic features of the elastic Hamiltonian (\ref{hamilt}) are

(i) the anisotropy in the harmonic part, as a consequence of which fluctuations
in the ``stiff'' directions are suppressed, and

(ii) the wave--vector dependence of the interaction (\ref{model1}) or
(\ref{model2}).

We note that the statics of model I can be mapped onto the Hamiltonian of the
uniaxial dipolar magnet using the transformation $S_{\bf k} = p Q_{\bf k}$
(Cowley 1976; Folk {\em et al.} 1977).

Terms which are irrelevant for the critical behaviour (in the renormalization
group sense) have already been omitted in the effective Hamiltonian
(\ref{hamilt}) and interactions (\ref{model1}),(\ref{model2}). As noted above,
there is in general more than one soft sector; however, the respective
interaction vanishes upon repeated application of the renormalization--group
transformation. Hence we may consider the different soft sectors as independent
and of the form (\ref{hamilt}).

The renormalization--group transformation appropriate for the anisotropic
Hamiltonian (\ref{hamilt}) consists, as usual, of two steps:
\medskip

(1) Eliminate wave vectors in the momentum shells $b^{-1} < p < 1$ for the
soft, and $b^{-2+\eta/2} < q < 1$ for the stiff sector, respectively.
\smallskip

(2) Rescale according to ${\bf p}' = b {\bf p}$,
${\bf q}' = b^{2-\eta/2} {\bf q}$, and $Q'_{\bf k'} = \zeta^{-1} Q_{\bf k}$.
\medskip

The elastic anisotropy is reflected in the different scaling for the soft and
stiff sectors. This procedure yields a new effective Hamiltonian with coupling
constants $r'$ and $u'$. The decisive transformation is the one for the
nonlinear coupling $u$,
\begin{equation}
u' = b^{4+m-2d} \left[ u - 36 u^2 C(r) \right] \quad , \label{utrans}
\end{equation}
where $C(r)$ stems from the one--loop bubble diagram; from Eq.~(\ref{utrans})
we can immediately read off the (upper) critical dimension as a function of the
dimensionality $m$ of the soft sector (Folk {\em et al.} 1976 {\em a,b})
\begin{equation}
d_c(m) = 2 + {m \over 2} \quad . \label{crtdim}
\end{equation}
For $d > d_c(m)$, the nonlinear coupling is irrelevant, and the system
approaches a Gaussian fixed point $u^* = 0$; its critical behaviour is thus
governed by the classical critical exponents. Only for $d < d_c(m)$ is a
non--trivial fixed point of the renormalization--group transformation
approached, and the exponents assume non--classical values.

The effective suppression of critical fluctuations now becomes evident. For
one--dimensional soft sector the critical dimension is $d_c(1) = 5/2$, and
consequently elastic phase transitions in three dimensions are characterized by
the classical critical exponents
\begin{equation}
m = 1: \quad \nu = \frac12 \, , \; \eta = 0 \, , \; \gamma = 1 \, , \;
   \beta = \frac12 \, , \; \alpha = 0 \, , \; \delta = 3 \quad . \label{claexp}
\end{equation}

In the case of a two--dimensional soft sector one has $d_c(2) = 3$, and one
thus finds classical behaviour with logarithmic corrections in three
dimensions. For instance, the static susceptibility (inverse elastic
coefficient) and the specific heat are given by
\begin{equation}
m = 2: \quad \chi \propto \tau^{-1} | \ln \tau |^{r_\chi} \, , \quad
                C \propto | \ln \tau |^{r_C} \quad , \label{lgcorr}
\end{equation}
where $\tau = | T - T_c | / T_c$. The exponents $r_\chi$ and $r_C$ for models
I and II are listed in table~\ref{logcor} (Folk {\em et al.} 1976 {\em a,b}).
For a second--order elastic transition in $d=3$ with $m \geq 2$, the local
fluctuations diverge: $\langle u^2 \rangle \propto | \log \tau |$ for $m=2$ and
$\langle u^2 \rangle \propto \tau^{-1/2}$ for $m=3$; hence the Debye--Waller
vanishes at $T_c$. It is conceivable that the actual phase transformation in
such systems will be of first order, not necessarily leading to the phase
indicated by the soft mode.

Elastic phase transitions with a two--dimensional soft sector occur, e.g., in
${\rm KCN}$ (Hauss\"uhl 1973), ${\rm NaCN}$ (Rowe {\em et al.} 1975), and
s-triazene (Smith \& Rae 1978; Rae 1978); in most of the cases, however, the
soft sector is one--dimensional.

\medskip
\centerline{\fbox{\parbox[b]{3cm}{Table \ref{logcor} near here}}}
\medskip

There seems to be no isotropic elastic system that would show the onset of a
shear instability, possibly as a precursor to melting. However, it is
conceivable that the case $m=d$ could apply to phase transitions in disordered
systems like polymer gels, which on a sufficiently large length scale can be
viewed as isotropic elastic media. For isotropic elastic transformations the
upper critical dimension would be $d_c(3) = 7/2$, and one would have
non--classical behaviour in three dimensions. For elastic phase transitions of
layers deposited on (preferrably amorphous) substrates, Eq.~(\ref{crtdim})
predicts non--classical behaviour.


\section{Dynamic properties}

\subsection{Soft acoustic phonons}

In order to derive the equations of motion for the soft acoustic phonons, we
construct an effective Lagrangean from the free energy (\ref{elasfe}) and a
kinetic term, reexpressed in terms of the displacement fields via
(\ref{strain}),
\begin{equation}
L[\{ u_i \}] = \int {\rho \over 2}
        \left( {\partial u_i({\bf x},t) \over \partial t} \right)^2 d^3x dt -
                                {\cal F}[\{ u_i \}] \quad ; \label{lagran}
\end{equation}
here, $\rho$ denotes the mass density of the unit cell. Following nonlinear
(generalized) elasticity theory the deterministic part of the equations of
motion can now be obtained by applying the variational principle
$\delta L[\{ u_i \}] / \delta u_i({\bf x},t) = 0$. Degrees of freedom other
than the displacement field will lead to damping and noise; thus our complete
Langevin--type equation of motion for the phonon normal modes becomes
(Folk {\em et al.} 1979)
\begin{equation}
M {\ddot Q}_{\bf k} = - {\delta H[\{ Q_{\bf k} \}] \over \delta Q_{\bf -k}}
        - M \Gamma_{\bf k} {\dot Q}_{\bf k} + r_{\bf k} \quad . \label{langeq}
\end{equation}
For acoustic phonons, the damping coefficient reads
\begin{equation}
\Gamma_{\bf k} = D p^2 + {\tilde D} q^2 \quad , \label{dampin}
\end{equation}
where the damping constants $D$ and ${\tilde D}$ for the soft and stiff
sectors, respectively, are different in general. The stochastic force
$r_{\bf k}$ results from the non--critical degrees of freedom; it has zero mean
and its fluctuations are related to the damping coefficient by an Einstein
relation
\begin{equation}
\langle r_{\bf k}(t) r_{{\bf k}'}(t') \rangle = 2 \Gamma_{\bf k} k_{\rm B} T
                \delta({\bf k}+{\bf k}') \delta(t-t') \quad , \label{einste}
\end{equation}
which guarantees that the equilibrium distribution is
\begin{equation}
{\cal P}[\{ P_{\bf k},Q_{\bf k} \}] \propto \exp \left[ - {1 \over k_{\rm B}T}
        \left( \sum_{\bf k} {| P_{\bf k} |^2 \over 2 M} + H[Q_{\bf k}] \right)
                                                \right] \quad , \label{eqdist}
\end{equation}
where $P_{\bf k}$ denotes the canonical momentum conjugate to $Q_{\bf k}$.

For $T > T_c$, the dynamic phonon susceptibility in mean--field approximation
follows immediately from the Fourier--transformed equation of motion
(\ref{langeq}) with (\ref{dampin}), and the harmonic part of the Hamiltonian
(\ref{hamilt}),
\begin{equation}
\chi({\bf k},\omega) = \left[ - M \omega^2 - i M \omega (D p^2 + {\tilde D}q^2)
                        + r p^2 + p^4 + q^2 \right]^{-1} \quad . \label{dysusc}
\end{equation}
The poles of $\chi({\bf k},\omega)$ determine the phonon dispersion relation.
Eq.~(\ref{dysusc}) thus shows that as $T \rightarrow T_c$
($r = \xi^{-2} \rightarrow 0$), the dispersion of the soft acoustic modes
(${\bf q} = 0$) turns from linear to quadratic; assuming that $D$ stays
constant, the soft mode finally becomes overdamped in the vicinity of $T_c$.
The dynamical phonon correlation function $D({\bf k},\omega)$, which is
directly measured in scattering experiments as part of the dynamic structure
factor $S({\bf k},\omega)$ (see \sect5$\,b\,$), is related to the dynamic
susceptibility via the (classical) fluctuation--dissipation theorem
\begin{equation}
D({\bf k},\omega) = {2 k_{\rm B} T \over \omega} {\rm Im} \chi({\bf k},\omega)
                                                        \quad ; \label{dycorr}
\end{equation}
it is depicted in Fig.~\ref{strfac} for ${\bf q} = 0$ and $M^{1/2} D = 1.0$,
both in the hydrodynamic regime ($p \xi_1 = 0.1$) and in the critical region
($p \xi_2 = \infty$).

\medskip
\centerline{\fbox{\parbox[b]{3cm}{Figure \ref{strfac} near here}}}
\medskip

Before turning to a more thorough analysis of the critical dynamics, we remark
that, to be precise, the adiabatic elastic constants should enter the dynamical
equations of motion, and not the isothermal elastic constants used in \sect2.
The general thermodynamic relation between adiabatic and isothermal elastic
constants (of second order) is (Leibfried \& Ludwig 1961)
\begin{equation}
c_{iklm}^{(2) {\rm ad}} - c_{iklm}^{(2) {\rm is}} =
        {T \over C_\varepsilon} \beta_{ik} \beta_{lm} \quad , \label{isoelc}
\end{equation}
where $\beta_{ik} = (\partial \sigma_{ik} / \partial T)_\varepsilon$ is the
temperature--derivative of the stress tensor and $C_\varepsilon$ is the
specific heat at constant strain. In certain symmetry directions the adiabatic
and isothermal sound velocities coincide; for instance, according to
(\ref{isoelc})
$c^{\rm ad}_{11}-c^{\rm ad}_{12} = c^{\rm is}_{11}-c^{\rm is}_{12}$ (switching
back to Voigt's notation). However, in a cubic crystal
$c^{\rm ad}_{11} = c^{\rm is}_{11} + T \beta^2 / C_\varepsilon$, and this
difference might be noticeable and important when data are actually fitted over
a wide temperature range.

\subsection{Critical dynamics}

In situations where the soft mode is a propagating hydrodynamic mode,
considerable information can be gained already by simply using the hydrodynamic
result in conjunction with dynamical scaling. Thus one finds for the sound
velocity
\begin{equation}
c_s = \sqrt{c/\rho} \propto \tau^{\gamma/2} \propto \xi^{-(2-\eta)/2}
                                                        \quad , \label{soundv}
\end{equation}
where $\rho$ is the mass density, $c \propto \tau^\gamma$ denotes an
appropriate combination of elastic constants, i.e., the inverse of the static
critical susceptibility, $\xi \propto \tau^{-\nu}$ is the correlation length, 
and the scaling relation $\gamma = \nu (2 - \eta)$ has been used. Therefore we
expect for the dynamical critical exponent $z = 2 - \eta / 2$.

Indeed, supplementing the static rescaling operations of \sect2$\,b\,$ by
$\omega' = b^z \omega$, and carrying out the dynamic renormalization group
program, to one--loop order the following recursion relations for the dynamical
parameters $M$ and $D$ are found (Folk {\em et al.} 1979):
\begin{eqnarray}
M'   &&= b^{4-\eta-2z} M \quad , \label{mtrans} \\
M'D' &&= b^{2-\eta-z} \left[ M D + u^2 E(D,M) \ln b \right]
                         \quad . \label{dtrans}
\end{eqnarray}
Here, the explicit form for $E(D,M)$ may be inferred from the one--loop 
integral, but is not important for what follows. At the Gaussian fixed point 
($u^* = 0$, hence $\eta = 0$) both relations lead to
\begin{equation}
z = 2 \quad , \label{dynexp}
\end{equation}
as expected. In the case of a one--dimensional soft sector, furthermore, the
validity of the mean--field results implies that
\begin{equation}
c_s \propto | T - T_c |^{1/2} \quad , \label{cscrit}
\end{equation}
while the damping coefficient is temperature--independent,
\begin{equation}
D \propto | T - T_c |^0 \quad ; \label{dpcrit}
\end{equation}
consequently the sound attenuation coefficient becomes
\begin{equation}
\alpha_s = {D \omega^2 \over 2 c_s^3} \propto | T - T_c |^{-3/2} \omega^2
                                                        \quad . \label{alcrit}
\end{equation}
For $m = 1$, the dynamical susceptibility is given by the mean--field
expression (\ref{dysusc}), see Fig.~\ref{strfac}. In the case of a
two--dimensional soft sector, $r = \xi^{-2}$ has to be replaced by the inverse
static susceptibility in Eq.~(\ref{lgcorr}). The power law (\ref{cscrit}) has
been confirmed by many experiments; the prediction for the phonon damping was
verified both by Brillouin scattering (Errandon\'ea 1981) and ultrasonic
attenuation experiments (Garland {\em et al.} 1984).

Although this is of no relevance in real crystals, it is interesting to note
that dynamical scaling breaks down in the hypothetical isotropic elastic system
for $d < d_c$. The origin of this peculiar behaviour is the appearance of a
dangerous irrelevant variable, namely $X = M^{1/2} D$, whose fixed--point value
diverges. Under these circumstances there is no definite dynamical exponent;
in the immediate vicinity of $T_c$ one finds $z = 2 + c \eta$ (as for the
relaxational model A), while in the hydrodynamic region the exponent for the
sound velocity is $z = 2 - \eta / 2$, but yet another value emerges for the
damping coefficient (Folk {\em et al.} 1979).


\section{Linear coupling of an optical to an acoustic phonon}

As mentioned in the Introduction, there are frequently situations where the
driving mechanism for the elastic instability is the softening of an optical
phonon which couples linearly to a transverse acoustic mode, e.g. in
${\rm KH_2PO_4}$ (${\rm KDP}$) and ${\rm LaP_5O_{14}}$. Then one has to study
the interaction of the acoustic and the optical modes. The ensuing coupled
system can be described alternatively by modes which are linear combinations of
these phonons; the actual soft mode is primarily acoustic and is governed by
the characteristic anisotropic elastic Hamiltonian. Hence the resulting
critical behaviour in three dimensions is classical, even if the optical modes
have their origin in short--range interactions that would by themselves allow
for non--classical critical fluctuations. In ${\rm KDP}$, there is a linear
coupling of the polarization to the shear $\varepsilon_{12}$
(Brody \& Cummins 1974). In ${\rm LaP_5O_{14}}$, a Raman--active optical phonon
coupled to $\varepsilon_{13}$ drives the transition (Fox {\em et al.} 1976).

A simplified Hamiltonian containing all the essential features is (Schwabl
1980)
\begin{eqnarray}
H[\{ q_{\bf k} \},\{ Q_{\bf k} \}] &&=
\frac12 \sum_{\bf k} \left[ \omega_0({\bf k})^2 \, | q_{\bf k} |^2 +
                        \omega_a({\bf k})^2 \, | Q_{\bf k} |^2 +
                g \, i k_x \, Q_{\bf k} q_{-{\bf k}} \right] \nonumber \\
&&\qquad + u \sum_{{\bf k}_1} \ldots \sum_{{\bf k}_4}
q_{{\bf k}_1} \ldots q_{{\bf k}_4} \, \Delta({\bf k}_1 + \ldots + {\bf k}_4)
                                                        \quad , \label{phcoup}
\end{eqnarray}
where $\Delta({\bf k}) = 1$ if ${\bf k}$ equals a reciprocal lattice vector
and zero otherwise. $Q_{\bf k}$ and $q_{\bf k}$ are the normal coordinates of
the acoustic and optical phonons, respectively, and their uncoupled dispersion
relations are
\begin{equation}
\omega_0({\bf k})^2 = a' (T - T_c^0) / T_c^0 +
                c_1 k_x^2 + c_2 k_y^2 + c_3 k_z^2 \quad , \label{optpho}
\end{equation}
and
\begin{equation}
\omega_a({\bf k})^2 = \alpha_1 k_x^2 + \alpha_2 k_y^2 + \alpha_3 k_z^2
                                                        \quad . \label{acupho}
\end{equation}
The third and fourth term in the Hamiltonian (\ref{phcoup}) represent the
linear coupling of the acoustic and the optical phonon, and the anharmonic
interactions of the optical phonons, respectively. Note that:

(i) The transition is driven by the optical phonon, Eq.~(\ref{optpho}). Due to
the short--range interactions, the optical phonon by itself would display
non--classical critical behaviour.

(ii) In ${\rm KDP}$ there are (uniaxial) dipolar forces present, and
$\omega_0({\bf k})^2$ thus contains terms $\propto k_z^2 / k^2$; hence, without
the coupling to the acoustic phonon, one would find logarithmic corrections to
the classical exponents in three dimensions.

(iii) Eqs.~(\ref{phcoup})--(\ref{acupho}) refer to the vicinity of the $k_x$
axis. If the acoustic phonon couples to $\varepsilon_{12}$, there is also a
contribution $\propto i k_y Q_{\bf k} q_{-{\bf k}}$ in Eq.~(\ref{phcoup}).

The dynamics of the coupled system are described by the stochastic equations of
motion
\begin{eqnarray}
{\ddot q}_{\bf k} &&=
- {\delta H[\{ q_{\bf k} \},\{ Q_{\bf k} \}] \over \delta q_{-{\bf k}}} - 
\Gamma_q({\bf k}) {\dot q}_{\bf k} + r_{q \, {\bf k}} \quad , \label{optdyn} \\
{\ddot Q}_{\bf k} &&=
- {\delta H[\{ q_{\bf k} \},\{ Q_{\bf k} \}] \over \delta Q_{-{\bf k}}} -
\Gamma_Q({\bf k}) {\dot Q}_{\bf k} + r_{Q \, {\bf k}} \quad . \label{acudyn}
\end{eqnarray}
The stochastic forces $r_{q \, {\bf k}}$ and $r_{Q \, {\bf k}}$ are related 
via Einstein relations of the form (\ref{einste}) to the damping coefficients,
which have the form $\lim_{{\bf k} \rightarrow 0} \Gamma_q({\bf k}) \not= 0$
and $\Gamma_Q({\bf k}) = D_1 k_x^2 + D_2 k_y^2 + D_3 k_z^2$. For small wave
vectors, the modes which diagonalize the harmonic part of the Langevin
equations of motion (\ref{optdyn}), (\ref{acudyn}) are given by
\begin{equation}
f_{1 \, {\bf k}} = {q_{\bf k} - \alpha({\bf k}) Q_{\bf k} \over
                \sqrt{1 + | \alpha({\bf k}) |^2}} \; , \quad
f_{2 \, {\bf k}} = {- \alpha({\bf k}) q_{\bf k} + Q_{\bf k} \over
                \sqrt{1 + | \alpha({\bf k}) |^2}} \quad , \label{diamod}
\end{equation}
with $\alpha({\bf k}) = i g k_x / 2 \omega_0(0)^2$, and the corresponding
eigenfrequencies are
\begin{equation}
\lambda_{1/2} = \omega_{0/a}({\bf k})^2 \pm {g^2 k_x^2 \over 4 \omega_0(0)^2}
                                                        \quad . \label{eigfre}
\end{equation}

For small wave vectors ${\bf k}$, the mode $f_{2 \, {\bf k}}$ is essentially
acoustic; its sound velocity in the $k_x$ direction, as obtained from
$\lambda_2$, vanishes at the temperature
\begin{equation}
T_c = T_c^0 + {T_c^0 g^2 \over 4 \alpha_1 a'} \quad . \label{critem}
\end{equation}
In terms of the new variables $f_{1/2 \, {\bf k}}$ the Hamiltonian reads
\begin{eqnarray}
&&H[\{ f_{1/2 \, {\bf k}}\}] =
\frac12 \sum_{\bf k} \left[ \lambda_1 \, | f_{1 \, {\bf k}} |^2 +
                \lambda_2 \, | f_{2 \, {\bf k}} |^2 \right] \label{diaham} \\
&&\qquad + u \sum_{{\bf k}_1} \ldots \sum_{{\bf k}_4}
        {f_{1 \, {\bf k}_1} + \alpha({\bf k_1}) f_{2 \, {\bf k}_1} \over
                                \sqrt{1 + | \alpha({\bf k}_1)|^2}} \ldots
        {f_{1 \, {\bf k}_4} + \alpha({\bf k_4}) f_{2 \, {\bf k}_4} \over
                                \sqrt{1 + | \alpha({\bf k}_4)|^2}} 
                \Delta({\bf k}_1 + \ldots + {\bf k}_4) \quad . \nonumber
\end{eqnarray}
It is important to realize that the terms involving $f_{2 \, {\bf k}}$ are
governed by the anisotropic Hamiltonian (\ref{hamilt}) with $m=1$ and the soft
direction $k_x$. Since $\lambda_1$ remains finite at $T_c$, $f_{1 \, {\bf k}}$
scales as $f_{1 \, {\bf k}} \propto b^{-2} f_{2 \, {\bf k}}$ and is thus
irrelevant for the critical behaviour. Therefore, the statics and dynamics of
the phase transition at $T_c$ are described by mean--field critical exponents
(Schwabl 1980), as previously suggested by a self--consistency argument
(Villain 1970).

Closing our discussion of pure crystals, we remark that interactions of the
soft acoustic phonon with other acoustic modes have been disregarded here. For
instance, terms of the form $\varepsilon_{ii} \times ({\rm shear})^2$ are
possible, which cause a coupling of the soft transverse mode to longitudinal
phonons, similar in structure to the magnetostrictive interaction in
compressible magnets (Wegner 1974; Bergman \& Halperin 1976). Eliminating the
non--critical longitudinal fluctuations leads to additional (negative)
fourth--order couplings betweeen the soft modes. One may thus anticipate that
in analogy to the magnetostrictive case, the transition becomes of first order
for free boundaries, and the specific--heat exponent will be positive,
$\alpha > 0$. For a clamped crystal, on the other hand, the transition would
remain continuous; however, clamping would interfere with the shear deformation
which is characteristic of the elastic systems under consideration here.


\section{Disorder effects}

\subsection{Local order parameter condensation}

Real crystals contain defects of all sorts: point disorder, dislocations, grain
boundaries, etc. The more complicated the unit cell, the more likely crystal
growth will lead to imperfections. Some of these defects will have a pronounced
impact on the phase transition. We thus turn to the investigation of the
influence of disorder on the statics and dynamics of elastic phase transitions.
More specifically, we shall be interested in a certain type of inhomogeneities,
which have the effect of locally changing the elastic constants, and thus
softening the crystal. In this Section, we shall illustrate some local features
by discussing an effectively one--dimensional model with a single impurity
(Schmidt \& Schwabl 1977, 1978; Schwabl \& T\"auber 1991 {\em a}). As a matter
of fact, three--dimensional systems behave qualitatively similar, when we
restrict ourselves to the soft sectors in momentum space. Understanding these
single--defect properties facilitates the interpretation of the results for
systems with a truly finite disorder concentration, to be addressed in
\sect5$\,b\,$ (Schwabl \& T\"auber 1991 {\em b}; Bulenda {\em et al.} 1995). 

The one--dimensional elastic model free energy reads (Schmidt \& Schwabl 1978)
\begin{equation}
{\cal F}[\varepsilon] = \int \left(
                        \frac12 \left[ a - U(x) \right] \varepsilon(x)^2 +
        {c \over 2} \varepsilon'(x)^2 + {b \over 4} \varepsilon(x)^4 \right) dx
                                                        \quad , \label{1delas}
\end{equation}
where the harmonic elastic constant vanishes at the transition temperature of
the pure crystal, $a = a' (T - T_c^0)$, and the Ginzburg--Landau parameters
$b$ and $c$ are taken to be positive constants. The disorder influence is
described by the short--range defect potential $U(x)$, which can be
approximated by a delta function, $U(x) = U_0 \delta(x)$, if both the
correlation length and the wavelengths of the characteristic fluctuations are
small compared to the typical defect size, which are assumed to be comparable
to the lattice constant $a_0$. An attractive potential with positive $U_0$
means that the transition temperature is locally increased. The probability
distribution ${\cal P}[\varepsilon]$ for a configuration $\varepsilon(x)$ is
given by
\begin{equation}
{\cal P}[\varepsilon] \propto \exp
    \left( - {\cal F}[\varepsilon] / k_{\rm B} T \right) \quad ; \label{1dprob}
\end{equation}
in the spirit of Ginzburg--Landau theory one determines the most probable state
${\bar \varepsilon}(x)$ from the stationarity condition
$\delta {\cal F}[\varepsilon] / \delta \varepsilon(x) = 0$; this
one--dimensional configuration then serves as an approximation for the real
three--dimensional order parameter (for a review of the application of Landau
theory to structural phase transitions, see, e.g., Salje, 1992).

Inserting (\ref{1dprob}) yields the nonlinear second--order differential
equation
\begin{equation}
c \varepsilon(x)'' = [a - U(x)] \varepsilon(x) + b \varepsilon(x)^3
                                                                \label{stateq}
\end{equation}
for the stationary state, which of course always has the constant solution
${\bar \varepsilon}_0 = 0$ (note, however, that the constant solutions for
$T < T_c^0$, ${\bar \varepsilon}_{\pm} = \pm \, \sqrt{|a| / b}$ describing the
two possible orientations of the ordered state, are no longer allowed if
$U_0 \not= 0$). Due to the defect influence this homogeneous state may become
unstable at a certain temperature $T_c^l > T_c^0$, and instead a localized
order parameter condensate (cluster) near the defect may form (Schmidt \&
Schwabl 1978); $T_c^l$ is called local transition temperature, although it does
not define a proper phase transition. Linear stability analysis shows that in
the one--dimensional model local condensation occurs at
\begin{equation}
T_c^l = T_c^0 + {U_0^2 \over 4 a' c} \quad , \label{loctrt}
\end{equation}
for any positive defect strength $U_0$. For $T < T_c^l$ one finds the following
stable cluster configurations (see Fig.~\ref{loccon}) in the interval
$T_c^0 \leq T \leq T_c^l$,
\begin{equation}
{\bar \varepsilon}_>(x) = {\pm \sqrt{2 a / b} \over
                        {\rm sinh}(|x|/\xi_> + \rho_>)} \quad , \label{opclgr}
\end{equation}
where $\xi_> = \sqrt{c / a}$ and $\rho_> = {\rm arcoth}(U_0 / 2 \sqrt{a c})$;
and in the temperature range $T < T_c^0$
\begin{equation}
{\bar \varepsilon}_<(x) = \pm \sqrt{|a| / b} \,
                        {\rm coth}(|x|/\xi_< + \rho_<) \quad , \label{opclsm}
\end{equation}
with $\xi_< = \sqrt{2 c / |a|}$ and
$\rho_< = (1/2) \, {\rm arsinh}(\sqrt{8 |a| c} / U_0)$. Note that the widths of
both configurations are given by the correlation lengths of the pure system
$\xi_>$ and $\xi_<$, respectively. We remark that in $d > 1$ dimensions, in
general a certain minimum disorder potential strength must be exceeded for
local condensation to occur (Schmidt \& Schwabl 1977; Schwabl \& T\"auber 1991
{\em b}); but the ensuing order parameter cluster still decays exponentially
$\propto e^{-r/\xi}$ for large $r \gg \xi$.

\medskip
\centerline{\fbox{\parbox[b]{3cm}{Figure \ref{loccon} near here}}}
\medskip

One can now look for traces of the local order parameter condensation in the
spectrum of the soft phonons. The one--dimensional model (\ref{1delas}) applies
for both distortive and elastic transitions, and hence the static
single--defect properties are completely analogous. However, their respective
dynamics are very different: In the case of distortive transitions, one may
obtain the spectrum of the optical phonons for $T \geq T_c^l$ by linearizing
Eq.~(\ref{stateq}) about the stationary solution ${\bar \varepsilon}_0 = 0$; as
a result of the ``attractive'' defect, one additional localized mode emerges
below the continuum of propagating scattering states, and its frequency
vanishes precisely at the local transition temperature (\ref{loctrt}). This
softening of a localized defect excitation can thus be viewed as a dynamical
precursor for the local condensation in distortive systems (Schmidt \& Schwabl
1978).

Similar localized acoustic phonon modes have been hypothesized for elastic
systems, namely for the related first--order martensitic transformations (Clapp
1973, 1979, 1981). When supplemented with a dynamical equation of the form
(\ref{langeq}), the simplified one--dimensional model (\ref{1delas}) permits an
analytic solution. Linearizing the equation of motion yields a fourth--order
differential equation for the soft acoustic modes, the eigenstates of which are
linear combinations of propogating ($\propto e^{\pm i k x}$) and exponentially
localized ($\propto e^{\pm \sqrt{\xi_>^{-2} + k^2} x}$) contributions.
Employing the appropriate boundary conditions for the delta--function defect,
one finds that there appears no additional mode in the elastic system; any
dispersion--free localized mode would be unstable towards decay into continuum
states. Instead, each of the propagating modes contains a localized vibrational
contribution in the defect vicinity, which for the long--wavelength modes
($k \rightarrow 0$) condenses as $T \rightarrow T_c^l$, in the sense that an
incoming phonon is not transmitted through the defect region, but the local
order parameter condensate is formed (Schwabl \& T\"auber 1991 {\em a}).
Accordingly, the delay time of an incoming wave packet diverges
$\propto (T - T_c^l)^{-1}$ as $T_c^l$ is approached, as does the relaxation
time of a stress--induced cluster above the local transition.

\subsection{Disorder--induced phase transition and central peak}

By appropriately averaging the single--defect solutions, one may already
infer some properties of a crystal with $N_{\rm D} \gg 1$ defects, the
concentration $n_{\rm D} = N_{\rm D}/N$ of which is to remain finite in the
thermodynamic limit (Schwabl \& T\"auber 1991 {\em a}). Instead, we proceed to
study such a system in $d$ space dimensions using a more direct approach
(Schwabl \& T\"auber 1991 {\em b}; Bulenda {\em et al.} 1995). It is now more
convenient to use a lattice representation, and we therefore define the
short--range quenched impurity potential produced by $N_{\rm D}$ point defects,
located on the randomly selected lattice sites $i_{\rm D}$ as
\begin{equation}
\phi_{ij} = \lambda \sum_{i_{\rm D}=1}^{N_{\rm D}}
                        \delta_{i,i_{\rm D}} \delta_{ij} \quad , \label{defpot}
\end{equation}
with Fourier transform (note that the system is not translation--invariant)
\begin{equation}
\phi_{{\bf k}{\bf k}'} = {1 \over N} \sum_{i,j=1}^N \phi_{ij}
        e^{-i({\bf k}{\bf x}_i - {\bf k}'{\bf x}_j)} \quad . \label{foudef}
\end{equation}
As in the previous Section, we require that $\lambda = U_0 / a_0^d > 0$ such
that the defects locally enhance the transition temperature. One can then write
the harmonic part of the elastic Hamiltonian (\ref{hamilt}), supplemented by
the disorder potential, as
\begin{equation}
H[\{ Q_{\bf k} \}] = \! \int \! d^dk \! \int \! d^dk' \frac12 \left[
                (a p^2 + \alpha q^2 + c p^4) \delta_{{\bf k}{\bf k}'} -
                        ({\bf k}{\bf k}') \phi_{{\bf k}{\bf k}'} \right]
        Q_{\bf k} Q_{-{\bf k}'} + {\cal O}(Q_{\bf k}^4) \; , \label{defham}
\end{equation}
where $a = a'(T - T_c^0)$, and $\alpha , b > 0$ are constants. In addition to
the thermal average, for this system with random quenched disorder each
physical quantity has to be averaged over all possible defect configurations;
the formal definition of this quenched disorder average is
\begin{equation}
\langle \langle \ldots \rangle \rangle = \prod_{j=1}^{N_{\rm D}} \left[
{1 \over N} \sum_{i_{{\rm D}_j}=1}^N \right] \ldots \quad . \label{quench}
\end{equation}

The soft--phonon dynamics is given by the Langevin equation (\ref{langeq}),
with Eqs.~(\ref{dampin}) and (\ref{einste}). In the pure system, the (``free'')
phonon propagator reads [see Eq.~(\ref{dysusc})]
\begin{equation}
\chi_0({\bf k},\omega)^{-1} =
- M \omega^2 - i M \omega (D p^2 + {\tilde D} q^2) + a p^2 + \alpha q^2 + c p^4
                                                        \quad . \label{frprop}
\end{equation}
Introducing an external field $h_{\bf k}$ in the equation of motion, inverting
its Fourier transform, taking the thermal average, and differentiating it with
respect to $h_{\bf k}$, one arrives at the following recursion relation for the
``full'' dynamic response function in the high--temperature phase,
\begin{equation}
\chi({\bf k},{\bf k}',\omega) = \chi_0({\bf k},\omega) \delta_{{\bf k}{\bf k}'}
        + \chi_0({\bf k},\omega) \sum_{{\bf k}''} ({\bf k}{\bf k}'')
\phi_{{\bf k}{\bf k}''} \chi({\bf k}'',{\bf k}',\omega) \quad ; \label{recrel}
\end{equation}
systematical iteration of Eq.~(\ref{recrel}), and then performing the
configurational average (\ref{quench}) finally yields the translationally
invariant response function $\chi({\bf k},\omega)$ (for a diagrammatic
representation, see Schwabl \& T\"auber 1991 {\em b}). Collecting all
contributions that are linear in the defect concentration $n_{\rm D}$
(single--site approximation), one eventually finds (Bulenda {\em et al.} 1995)
\begin{equation}
\chi({\bf k},\omega)^{-1} = \chi_0({\bf k},\omega)^{-1} -
{\lambda n_{\rm D} (p^2 + q^2) \over 1 - \lambda (a_0 / 2 \pi)^d \int^\Lambda
(p^2 + q^2) \chi_0({\bf k},\omega) d^mp d^{d-m}q} \quad , \label{resfun}
\end{equation}
where the momentum cutoff $\Lambda \approx 2 \pi / a_0$ has been introduced.

The result (\ref{resfun}) implies that as a consequence of the coupling to the
softening defects, the entire system may become unstable towards a new
low--temperature phase with finite average order parameter at a certain
temperature $T_c(n_{\rm D})$, which is to be determined from the condition
\begin{equation}
\lim_{k \rightarrow 0} \left[ k^{-2} \chi({\bf k},\omega=0)^{-1} \right] = 0
                                                        \quad . \label{instab}
\end{equation}
As for distortive transitions (Schwabl \& T\"auber 1991 {\em b}), for $d > 1$ a
certain minimum defect strength is required for this instability to occur, and
$T_c(n_{\rm D})$ is bounded below by the local transition temperature $T_c^l$
of \sect5$\,a\,$, which can be considerably higher than $T_c^0$, the transition
temperature of the pure system.

In the response and correlation functions, the singularity at $T_c(n_{\rm D})$
appears as a dynamical central peak. In view of the single--defect properties
of the previous Section, one may understand this peak as stemming from
overlapping localized vibrations at the point defects which ``condense'' at
$T_c(n_{\rm D})$, thereby forming a spatially inhomogeneous order parameter
configuration. In Fig.~\ref{phocor} the dynamical correlation function
(\ref{dycorr}) is depicted for the case of an elastic system with a
one--dimensional soft sector; the numerical values used for the
Ginzburg--Landau parameters there are appropriate for ${\rm Nb_3Sn}$
($T_c^0 = 45 {\rm K}$), and the defect strength $\lambda$ was adjusted
arbitrarily in order that $T_c(n_{\rm D}) = 65 {\rm K}$
(Bulenda {\em et al.} 1995). As $T \rightarrow T_c(n_{\rm D})$ from above, the
soft phonon peak at finite energy is shifted to lower frequencies, and a very
sharp and distinctive dynamical central peak emerges [Fig.~\ref{phocor}(a)].
Fig.~\ref{phocor}(b) shows the dependence on the angle $\theta$ between the
external wave vector ${\bf k}$ and the soft sector; both graphs demonstrate
that the dynamical central peak is confined to temperatures very close to the
defect--induced phase transition at $T_c(n_{\rm D})$, and to wave vectors
within the soft sector, reflecting the fact that wave vectors in the stiff
directions do not probe the critical properties of the crystal.

\medskip
\centerline{\fbox{\parbox[b]{3cm}{Figure \ref{phocor} near here}}}
\medskip

The properties of the ensuing inhomogeneous low--temperature phase may be
studied using a self--consistent mean--field approach (Schwabl \& T\"auber
1991 {\em b}; Bulenda {\em et al.} 1995). The starting point is the following
discrete version of the nonlinear Ginzburg--Landau free energy for a
single--component order parameter $\varepsilon_i$ in $d$ dimensions
[cf. Eqs.~(\ref{elasfe}) and (\ref{1delas})],
\begin{equation}
{\cal F}[\varepsilon_i] =
\frac12 \sum_{i,j=1}^N \varepsilon_i G_{0 \, ij}^{-1} \varepsilon_j -
{\lambda \over 2} \sum_{i=1}^N \sum_{i_{\rm D}=1}^{N_{\rm D}}
                                        \varepsilon_i^2 \delta_{i,i_{\rm D}} +
      {b \over 4} \sum_{i=1}^N \varepsilon_i^4 - \sum_{i=1}^N h_i \varepsilon_i
                                                        \quad , \label{discfe}
\end{equation}
where a stress term $\propto h_i$ has been introduced, and the static
propagator $G_{0 \, ij}^{-1}$ is defined by its Fourier transform
\begin{equation}
G_0({\bf k})^{-1} = {a p^2 + \alpha q^2 + c p^4 \over k^2}
                                                        \quad . \label{stprop}
\end{equation}
In the framework of the Ginzburg--Landau approximation, i.e., neglecting order
parameter fluctuations, the stationarity condition becomes (with constant
external stress $h_i = h$)
\begin{equation}
\sum_{j} G_{0 \, ij}^{-1} \varepsilon_j -
\lambda \sum_{i_{\rm D}} \varepsilon_i \delta_{i,i_{\rm D}} + b \varepsilon_i^3
                                                = h \quad . \label{stcond}
\end{equation}

The solution of Eq.~(\ref{stcond}), with its combined nonlinearity and
randomness, in general poses a formidable problem. Therefore an additional
approximation is employed, namely the following ansatz for the thermodynamical
average ${\bar \varepsilon}_i$ of the order parameter is used,
\begin{equation}
{\bar \varepsilon}_i = A + B \sum_{i_{\rm D}} \delta_{i,i_{\rm D}}
                                                        \quad , \label{ansatz}
\end{equation}
i.e.: the order parameter at each lattice point $i$ is assumed to be the sum of
a homogeneous background $A$ and an additional contribution $B$, if there is a
defect at site $i$, thus enhancing the total value of the order parameter to
$A + B$ (see Fig.~\ref{opconf}). Thus we use the approximation that at all
defect sites the order parameter points in the same direction, and in addition
neglect the spatial variation of the order parameter near the defects. This
seemingly rather crude ansatz already contains the possible relevant
modifications which can be caused by the impurities, namely (i) an enhancement
of the spatially averaged order parameter (corresponding to the parameter $A$),
and (ii) the ensuing ``screening'' of the defect potential (described by $B$).

\medskip
\centerline{\fbox{\parbox[b]{3cm}{Figure \ref{opconf} near here}}}
\medskip

Inserting the ansatz (\ref{ansatz}) into the stationarity equation
(\ref{stcond}) yields
\begin{equation}
{\bar \varepsilon}_{\bf k} = h \delta_{{\bf k},{\bf 0}} {\tilde G}_0({\bf k}) +
{\tilde G}_0({\bf k}) \sum_{{\bf k}'} \tilde{\phi}_{{\bf k}{\bf k}'}
                        {\bar \varepsilon}_{{\bf k}'} \quad , \label{oprecr}
\end{equation}
where a renormalized propagator
\begin{equation}
{\tilde G}_0({\bf k})^{-1} = G_0({\bf k})^{-1} + b A^2 \label{renpro}
\end{equation}
and screened defect potential $\tilde{\phi}_{{\bf k}{\bf k}'}$ with weakened
strength
\begin{equation}
{\tilde \lambda} = \lambda - b \left[ (A+B)^2-A^2 \right] \label{rendis}
\end{equation}
have been introduced. As Eq.~(\ref{recrel}) for the dynamics in the
high--temperature phase, the recursion relation (\ref{oprecr}) can be iterated
and then the quenched disorder average performed, with the single--site
approximation result
\begin{equation}
\langle \langle {\bar \varepsilon} \rangle \rangle \left[ a + b A^2 -
{{\tilde \lambda} n_{\rm D} \over 1 - {\tilde \lambda} (a_0 / 2 \pi)^d
\int_0^\Lambda {\tilde G}_0({\bf k}) d^dk} \right] = h \quad ; \label{opdet1}
\end{equation}
on the other hand, immediate averaging of Eq.~(\ref{stcond}) gives
\begin{equation}
(a + b A^2) (A + n_{\rm D} B) - {\tilde \lambda} n_{\rm D} (A + B) = h
                                                        \quad . \label{opdet2}
\end{equation}
Eqs.~(\ref{opdet1}) and (\ref{opdet2}) constitute two coupled nonlinear
equations that uniquely determine the mean order parameter; assuringly, for
$h \rightarrow 0$ they yield non--zero self--consistent solutions for
$\langle \langle \varepsilon \rangle \rangle$ precisely when
$T < T_c(n_{\rm D})$, the transition point determined from the instability of
the high--temperature phase. Fig.~\ref{meanop} shows that the order parameter
of the disordered crystal as function of $T$ looks similar to the corresponding
curve for the pure system, with the singularity at $T_c^0$ being smeared out by
the defects (in the graph, the reduced temperature $t = (T - T_c^0) / T_c^0$ is
used). The order parameter sets in continuously at $T_c(n_{\rm D})$, with the
usual mean--field exponent $\beta = 1/2$, remains minute in the range
$T_c(n_{\rm D}) > T > T_c^0$, and starts to grow to larger values only near
$T_c^0$. Thus the transition temperature of the pure system remains an
important parameter even in the perturbed system, although the true phase
transition takes place at $T_c(n_{\rm D})$, which, however, may in fact be
hardly noticeable in experiments. Again, the results for a three--dimensional
system with one one--dimensional soft sector are depicted, but the qualitative
features are essentially the same in the cases of a two-- (or even
three--)dimensional soft sector.

\medskip
\centerline{\fbox{\parbox[b]{3cm}{Figure \ref{meanop} near here}}}
\medskip

Inserting the self--consistent solutions for the mean order parameter as
function of temperature into the free energy (\ref{discfe}), the specific heat
in Landau approximation is readily determined by taking the derivatives with
respect to $T$,
$C_v = - T (\partial^2 {\cal F}[{\bar \varepsilon}] / \partial T^2)_V$; see
Fig.~\ref{spheat}. Obviously, the discontinuity at $T_c^0$ has been smeared
out, in place of which a tiny jump emerges at $T_c(n_{\rm D})$. Note that there
is a very distinct maximum of the specific heat near $T_c^0$, while the
extremely minute jump at $T_c(n_{\rm D})$ might not be experimentally
detectable at all.

\medskip
\centerline{\fbox{\parbox[b]{3.5cm}{Figure \ref{spheat} near here}}}
\medskip

The dynamical phonon response and correlation functions may similarly be
calculated following the above procedures; using the ansatz (\ref{ansatz})
leads to a recursion relation precisely of the form (\ref{recrel}), where
modified parameters
\begin{equation}
a \rightarrow a + 3 b A^2 \; , \quad \alpha \rightarrow \alpha + 3 b A^2 \, ,
        \quad \lambda \rightarrow \lambda - 3 b B (2 A + B) \label{dynmod}
\end{equation}
are to be inserted. One finds that the dynamical central peak in the phonon
correlation function rapidly disappears as the temperature is lowered below
$T_c(n_{\rm D})$.

However, there is a very interesting additional static contribution to the
dynamic structure factor, caused by the defect--induced spatial inhomogeneity
of the order parameter. The dynamic structure factor, as measured in scattering
experiments, is defined as
\begin{equation}
S({\bf k},\omega) = \int e^{i \omega t} \Bigg \langle {1 \over N}
        \sum_{i,j=1}^N e^{-i {\bf k} [{\bf a}_i + {\bf u}_i(t)]} \,
                e^{i {\bf k} [{\bf a}_j + {\bf u}_j(0) ]} \Bigg \rangle dt
                                                        \quad , \label{factor}
\end{equation}
where ${\bf a}_i$ denote the Bravais lattice sites, and ${\bf u}_i$ the
displacements from these equilibrium positions. In order to evaluate
$\langle \langle S({\bf k},\omega) \rangle \rangle$, we perform a cumulant
expansion for the combined thermal and configurational averages of the
exponentials in (\ref{factor}) and expand to second order, decomposing the
displacement fields ${\bf u}_i(t)$ into a static part ${\bar{\bf u}}_i$ and a
fluctuating contribution ${\tilde{\bf u}}_i(t)$. Eventually one arrives at
(Schwabl \& T\"auber 1991 {\em b}; Bulenda {\em et al.} 1995)
\begin{eqnarray}
\langle \langle S({\bf k},\omega) \rangle \rangle &&= 2\pi \delta(\omega)
        \left[ N \sum_{\bf g} \delta_{{\bf k},{\bf g}} + \sum_{\alpha \beta}
k^\alpha k^\beta \langle \langle S_c^{\alpha \beta}({\bf k}) \rangle \rangle
                                                \right] e^{-2W} \nonumber \\
&&\qquad + \left[ \sum_{\alpha \beta} k^\alpha k^\beta
        D^{\alpha \beta}({\bf k},\omega) \right] e^{-2W} \quad , \label{meanst}
\end{eqnarray}
where the first term represents the elastic Bragg scattering peaks occuring at
the reciprocal lattice vectors ${\bf g}$, $W$ is the Debye--Waller factor, and
\begin{equation}
S_c^{\alpha \beta}({\bf k}) = {1 \over N} \sum_{i,j}
e^{-i {\bf k} ({\bf a}_i-{\bf a}_j)} \left( {\bar u}_i^\alpha {\bar u}_j^\beta
        - \langle \langle {\bar u}^\alpha \rangle \rangle
        \langle \langle {\bar u}^\beta \rangle \rangle \right) \label{huangs}
\end{equation}
yields an additional static contribution to the structure factor arising from
elastic scattering from random variations of the local order parameter (Huang
scattering); finally, $D^{\alpha \beta}({\bf k},\omega)$ is the dynamical
phonon correlation function (\ref{dycorr}) and (\ref{resfun}), discussed
before, describing inelastic scattering processes.
 
In the framework of the above self--consistent mean--field theory and
single--site approximation, the Huang scattering amplitude is readily obtained
observing that ${\bar \epsilon}_{\bf k} = {\bf k}{\bar{\bf u}}_{\bf k}$, and
using iterations of Eq.~(\ref{oprecr}), with the result
\begin{equation}
k^2 S_c^{\alpha \beta}({\bf k}) = {n_{\rm D} {\tilde \lambda}^2
        \langle \langle {\bar \varepsilon} \rangle \rangle^2 k^\alpha k^\beta
       {\tilde G}_0({\bf k})^2 \over \left[1 - {\tilde \lambda} (a_0 / 2 \pi)^d
                        \int^\Lambda {\tilde G}_0({\bf k'}) d^dk' \right]^2} =
      (a + b A^2)^2 {\langle \langle {\bar \varepsilon} \rangle \rangle^2 \over
     n_{\rm D}} k^\alpha k^\beta {\tilde G}_0({\bf k})^2 \quad , \label{huares}
\end{equation}
where in the final step the condition (\ref{opdet1}) (for $h \rightarrow 0$)
was used. Note that the renormalized static susceptibility, which determines
this additional elastic scattering amplitude, does not diverge at $T_c^0$;
furthermore, this static central peak is characterized by a finite width in
momentum space, $\gamma = \sqrt{(a + b A^2) / c}$ (if ${\bf k}$ lies in the
soft sector), which vanishes at the defect--induced transition temperature
$T_c(n_{\rm D})$. Fig.~\ref{huasca} depicts the Huang scattering amplitude
(\ref{huares}) as function of temperature for different wave vectors
${\bf k} = {\bf p}$ in the soft sector. The static central peak emerges at
$T_c(n_{\rm D})$, and grows to considerable intensity near $T_c^0$, but becomes
less prominent and widens in ${\bf k}$ space for $T < T_c^0$, as the order
parameter slowly approaches the spatially homogeneous configuration of the
undisturbed system. The dynamical central peak found in the phonon correlation
function near $T_c(n_{\rm D})$ may be viewed as its dynamical precursor in the
critical region, as well as the static one due to the new Bragg peaks which of
course persist through the entire low--temperature phase.

\medskip
\centerline{\fbox{\parbox[b]{3.5cm}{Figure \ref{huasca} near here}}}
\medskip

At this point we have to comment on the applicability of our mean--field
approach. Although we have explained in \sect2 that critical fluctuations
only play a minor role for elastic phase transitions, the situation here is
somewhat different, for we now have to worry about local fluctuations in the
orientations of the disorder--induced order parameter clusters (\ref{opclgr}),
which were entirely neglected in the above treatment. The mean--field
approximation suggests that as soon these local condensates form at
$T_c(n_{\rm D}) \approx T_c^l$, a coherent order parameter with non--zero mean
emerges, independent of space dimension $d$. More realistically, one would
expect that while at $T_c^l$ local condensation at the defects takes place, yet
the orientations of the different clusters still fluctuate considerably, and
they possibly form a true new ground state only at a lower temperature
$T_{\rm ord}$. A crude estimate of this true ordering temperature is obtained
by the criterion that the correlation length should at least be of the order of
the mean defect separation $\propto n_{\rm D}^{-1/d}$ for collective behaviour
of the condensates to occur. A more refined argument considers the different
cluster orientations as effectively Ising--like degrees of freedom, and
determines $T_{\rm ord} = J / k_{\rm B}$, where $J$ is the free--energy
difference between configurations with parallel and opposite cluster
orientations (Bulenda {\em et al.} 1995). In $d=3$ dimensions, the resulting
``true'' phase transition temperature is considerably lower than
$T_c(n_{\rm D})$ for identical disorder concentration, but may still be well
above $T_c^0$; thus, at least for $T \leq T_{\rm ord}$ a spatially
inhomogeneous order parameter appears, which produces a static central peak
with finite width in momentum space for temperatures above the transition
temperature of the pure system, where the average order parameter is still very
small. In addition, the time scale of the cluster orientation fluctuations will
diverge $\propto (T - T_{\rm ord})^{-1}$ upon approaching $T_{\rm ord}$, which
in experiment would eventually render these indistinguishable form static
inhomogeneities. Qualitatively, at least, the situation will then appear as
shown in Figs.~\ref{meanop}--\ref{huasca}, with $T_c(n_{\rm D})$ replaced by
$T_{\rm ord}$ (or even a somewhat higher temperature, depending on the
experimental frequency resolution).

\subsection{Disorder and critical properties}

Another important issue is the question if disorder may change the critical
properties near elastic phase transitions, i.e., if the static and dynamical
critical exponents are determined by a new renormalization--group fixed point.
We briefly discuss the two relevant cases of (i) random--fields, and (ii)
random--$T_c$ disorder, and their possible influence in the critical region
very close to $T_c$ (Morgenstern 1988).

Quenched random fields couple linearly to the order parameter, and are taken to
have vanishing average $\langle h_{\bf k} \rangle$ and second moment
\begin{equation}
\langle h_{\bf k} h_{\bf k'} \rangle =
        \Delta k^{-\Theta} \delta({\bf k}-{\bf k'}) \quad , \label{ranfld}
\end{equation}
where $\Theta = 0$ for short--range disorder correlation, while
$0 < \Theta < m$ for long--range correlated defects. Analyzing the
renormalization--group recursion relations, one finds that these random fields
may alter the pure result for the (upper) critical dimension (\ref{crtdim}) to
\begin{equation}
d_c(m) = 3 + {m \over 2} + {\Theta \over 2} \quad , \label{rfcdim}
\end{equation}
where $m$ denotes the dimension of the soft sector (Morgenstern 1988). One
therefore expects non--classical critical exponents in three--dimensional
crystals undergoing an elastic phase transformation under the influence of
random strain fields.

For disorder coupling quadratically to the order parameter, i.e., of the
random--$T_c$ type, on the other hand, a variant of the Harris criterion may be
formulated. Again introducing possible long--range correlations
\begin{equation}
\langle \delta r_{\bf k} \delta r_{\bf k'} \rangle =
        W k^{-\Theta} \delta({\bf k}-{\bf k'}) \quad , \label{randtc}
\end{equation}
one finds that the second--order elastic phase transitions remains to be
governed by the pure critical exponents, provided that
\begin{equation}
\alpha + \nu \Theta < 0 \quad , \label{harris}
\end{equation}
where $\alpha$ and $\nu$ denote the specific heat and correlation length
critical exponents, respectively (Morgenstern 1988). For the case of point
defects ($\Theta = 0$) this means that new randomness--induced critical
exponents emerge if $\alpha > 0$, which is thus only possible for Ising--type
systems.


\section{Summary and comparison with related materials}

In conclusion, we summarize the above results and contrast them with related
displacive structural phase transformations, namely (1) second--order
distortive transitions, and (2) first--order martensitic transitions.

For second--order distortive structural phase transitions, the order parameter
is the displacement field corresponding to a soft optical phonon, while for
elastic transformations the relevant collective variable is a certain
combination of strain tensor components, and the soft mode is typically a
transverse acoustic phonon. This means that while in the distortive case at
$T_c$ one of the optical modes softens at a specific ${\bf k}$ vector, an
elastic instability implies the vanishing of the sound velocity along the soft
sector in the Brillouin zone. As we have seen in \sect2, as a consequence of
the ensuing strong anisotropy critical fluctuations are suppressed, and
continuous elastic phase transitions in three dimensions are governed by the
classical critical exponents (for a one--dimensional soft sector) with at most
logarithmic corrections to the scaling functions (for a two--dimensional soft
sector). For distortive structural transitions, on the other hand, critical
fluctuations can be quite prominent and have important effects on the values of
the critical exponents (see, e.g., Bruce \& Cowley 1981). Accordingly, the
dynamical critical exponent is $z = 2$ for elastic transitions (with $m < d$),
as in mean--field theory (\sect3), in contrast to the distortive case where
typically $z = 2 - c \eta$ with $\eta > 0$. Elastic instabilities may actually
be driven by the softening of an optical mode which is linearly coupled to the
transverse acoustic phonons; but even in that situation the ensuing critical
behaviour turns out to be classical; nevertheless, an anomaly is found in the
sound attenuation coefficient (\sect4).

We remark that the anisotropic Hamiltonian (\ref{hamilt}) applies also to
spin reorientation transitions (Hornreich \& Shtrikman 1976), to the phason
instability at $T = 49 {\rm K}$ in TTF-TCNQ (Bak 1976), and to the transition
from smectic-A to smectic-C in a magnetic field (Hornreich \& Shtrikman 1977). 
Anisotropic interactions are also present at Lifshitz points
(Hornreich {\em et al.} 1975).

The phenomenon of local order parameter condensation, induced by defects which
locally increase the transition temperature, is qualitatively similar for both
distortive and elastic phase transitions. However, dynamically the mechanism of
condensation is quite different; while in the distortive case there appear
localized modes below the continuum, which then soften at the local transition
temperature $T_c^l$, there appears no such additional localized mode in the
elastic case. Instead, the disorder leads to localized vibrational components
in the propagating modes, which then ``condense'' at $T_c^l$ to form the
defect--induced order parameter clusters (\sect5$\,a\,$). As was explained in
\sect5$\,b\,$, in a system with finite disorder concentration $n_{\rm D}$ a
true defect--induced phase transition can occur at $T_c(n_{\rm D})$, resulting
in a low--temperature state characterized by an initially strong spatial
inhomgeneity, which finally becomes smoothened out at low temperatures, where
the disorder influence is suppressed (\sect5$\,b\,$). A dynamical precursor for
this singularity at $T_c(n_{\rm D})$ (or $T_{\rm ord}$) is the appearance of a
dynamical central peak in the critical region, which only in the case of
distortive transitions can be traced back to the softening of a separate phonon
impurity band. In the low--temperature phase, besides the new Bragg peaks an
additional elastic Huang scattering component emerges as a consequence of the
spatial variations of the local order parameter in either case. Near the
transition temperature $T_c^0$ of the pure system, thermodynamic quantities
like the order parameter susceptibility and the specific heat appear
characteristically rounded. Both these features have been observed in a variety
of experiments.

As was discussed in \sect2, elastic instabilites often lead to first--order
phase transitions, as is the case for the interesting class of martensitic
transformations. These materials display a considerably more complex behaviour
than the second--order transitions we have discussed here (the precursor
effects in martensitic materials are reviewed in Finlayson, 1983). However, a
simplifying elastic Ginzburg--Landau free energy has been proposed (Falk 1980,
1983), which is of the form (\ref{elasfe}), but with a negative fourth--order
term and including a sixth--order term in order to describe a first--order
transition. Recently, an explanation for the experimentally observed ``tweed''
pattern in microscopic images of martensitic materials for temperatures far
above the transition temperature has been suggested on a similar basis (Kartha
{\em et al.} 1991, Sethna {\em et al.} 1992, Kartha {\em et al.} 1995). Namely,
such pseudo--periodic lattice deformations were found to emerge in an elastic
model including defects that locally modify the transition temperature in a
random manner. This intrinsic compositional disorder was found to conspire with
the natural geometric constraints of the lattice to form a frustrated, glassy
``tweed'' phase; and the random order parameter orientations then produce a
static central peak with finite width in momentum space. We finally note that
an interesting and wide field not covered in the present brief review concerns
orientational glasses found in mixed--crystal solids 
(H\"ochli {\em et al.} 1990).

Solid state physics in recent years has gone beyond the mere study of the
materials offered by our surrounding nature, but more and more aims at the
creation of new materials with taylored properties. As in semiconductor
physics and magnetism, a similar development is about to take place in
ferroelastic and ferroelectric substances. Besides of in--depth studies of
disorder, the investigation of artificial multilayers and other composite
structures will be of importance in the future.


\begin{acknowledgments}

U.C.T. acknowlegdes support from the Deutsche Forschungsgemeinschaft (DFG)
under Contract Ta.~177/1-2.

\end{acknowledgments}

\newpage


\newpage


\begin{table}
 \caption{Elastic phase transitions.}
 \label{elptra}
 \begin{tabular}{ccccc}
  \hline
  High--temperature & Vanishing combination & Strain & $m$ & Third--order \\
       phase        & of elastic constants  &        &     & invariants \\
  \hline
  Orthorhombic  & $c_{44}$ & $\varepsilon_{23}$ & $1$ & --- \\
                & $c_{55}$ & $\varepsilon_{13}$ & $1$ & --- \\
                & $c_{66}$ & $\varepsilon_{12}$ & $1$ & --- \\
  Tetragonal II & $c_{44}$ & $\varepsilon_{23}$, $\varepsilon_{13}$ & $1 + 2$ &
                                                                         --- \\
  Tetragonal I  & $c_{44}$ & $\varepsilon_{23}$, $\varepsilon_{13}$ & $1 + 2$ &
                                                                         --- \\
                & $c_{66}$ & $\varepsilon_{12}$ & $1$ & --- \\
         & $c_{11}-c_{12}$ & $\varepsilon_{11}-\varepsilon_{22}$ & $1$ & --- \\
  Cubic II & $c_{44}$ & $\varepsilon_{12}$, $\varepsilon_{13}$,
                                                     $\varepsilon_{23}$ & $2$ &
                  $\varepsilon_{23} \, \varepsilon_{13} \, \varepsilon_{12}$ \\
           & $c_{11}-c_{12}$ & $e_3$,$e_2$ & $1$ & $e_3 (e_3^2 - 3 e_2^2)$, \\
                                           & & & & $e_2 (e_3^2 - 3 e_2^2)$ \\
  Cubic I  & $c_{44}$ & $\varepsilon_{12}$, $\varepsilon_{13}$,
                                                     $\varepsilon_{23}$ & $2$ &
                  $\varepsilon_{23} \, \varepsilon_{13} \, \varepsilon_{12}$ \\
           & $c_{11}-c_{12}$ & $e_3$,$e_2$ & $1$ & $e_3 (e_3^2 - 3 e_2^2)$ \\
  Hexagonal II & $c_{44}$ & $\varepsilon_{23}$, $\varepsilon_{13}$ & $1 + 2$ &
                                                                         --- \\
           & $c_{66} = \frac12 (c_{11}-c_{12})$ &
                $\varepsilon_{12}$, $\varepsilon_{11}-\varepsilon_{22}$ & $2$ &
                $(\varepsilon_{11}-\varepsilon_{22})^3$, \\
           & & & & $\varepsilon_{12} (\varepsilon_{11}-\varepsilon_{22})^2$, 
                   $\varepsilon_{12}^3$ \\
  Hexagonal I  & $c_{44}$ & $\varepsilon_{23}$, $\varepsilon_{13}$ & $1 + 2$ &
                                                                         --- \\
           & $c_{66} = \frac12 (c_{11}-c_{12})$ &
                $\varepsilon_{12}$, $\varepsilon_{11}-\varepsilon_{22}$ & $2$ &
                $(\varepsilon_{11}-\varepsilon_{22})^3$, \\ 
           & & & & $\varepsilon_{12} (\varepsilon_{11}-\varepsilon_{22})^2$ \\
  \hline
 \end{tabular}
\end{table}

\begin{table}
 \caption{Logarithmic corrections.}
 \label{logcor}
 \begin{tabular}{ccc}
  \hline
  Model & $r_\chi$ & $r_C$ \\
  \hline
  I     &  $1/3$   & $1/3$ \\
  II    &  $4/9$   & $1/9$ \\
  \hline
 \end{tabular}
\end{table}


\begin{figure}
 \caption{Temperature dependence of the lattice parameters (a) and the
          monoclinic angle $\beta$ (b) in ${\rm NaOH}$
          (from Bleif {\em et al.} 1971).}
 \label{elcons}
\end{figure}

\begin{figure}
 \caption{Elastic moduli $c_{11}$, $c_{12}$, and $B$ of ${\rm Nb_3Sn}$ vs.
          temperature (from Rehwald {\em et al.} 1972).}
 \label{sndvel}
\end{figure}

\begin{figure}
 \caption{(a) Wave vectors and polarization vectors of soft transverse phonons
              in orthorhombic crystals; the sound velocities $c_s$ are:
              $(c_{44}/\rho)^{1/2}$ (dashed), $(c_{55}/\rho)^{1/2}$ (solid),
              $(c_{66}/\rho)^{1/2}$ (dotted).
          (b) and (c) Wave vectors and polarization vectors of soft elastic
              modes in tetragonal crystals.}
 \label{orttet}
\end{figure}

\begin{figure}
 \caption{(a) and (b) Elastic soft modes in cubic crystals, with (a) wave
              vector in one of the base planes and polarization perpendicular
              to it, $c_s = (c_{44}/\rho)^{1/2}$, and (b) wave vector in one of
              the face diagonals, $c_s = [(c_{11}-c_{12})/2\rho]^{1/2}$;
          (c) Soft transverse phonons in the hexagonal system, with sound
              velocities $(c_{44}/\rho)^{1/2}$ (dashed) and
              $[(c_{11}-c_{12})/2\rho]^{1/2}$ (solid).}
 \label{cubhex}
\end{figure}

\begin{figure}
 \caption{Dynamical correlation function for the soft acoustic phonon vs.
          $\omega M^{1/2}$ for $q = 0$, $M^{1/2} D = 1.0$, and two values of
          the correlation length $\xi$; solid line: $p \xi_1 = 0.1$
          (hydrodynamic regime), dashed line: $p \xi_2 = \infty$ (critical
          region).}
 \label{strfac}
\end{figure}

\begin{figure}
 \caption{Local condensation: Scetch of the order parameter profiles in a
          single--defect system in the temperature ranges $T < T_c^0$,
          $T_c^0 \leq T \leq T_c^l$, and $T > T_c^l$, respectively.}
 \label{loccon}
\end{figure}

\begin{figure}
 \caption{Dynamical phonon correlation function (with model parameters
          appropriate for ${\rm Nb_3Sn}$):
          (a) for different temperatures $T$ and fixed angle $\theta = 0$, and
          (b) for fixed temperature $T = 65.01 {\rm K}$ with different angles
              $\theta$ between external wave vector and soft sector
              ($T_c(n_{\rm D}) = 65 {\rm K}$).}
 \label{phocor}
\end{figure}

\begin{figure}
 \caption{Schematic illustration of the order parameter configuration according
          to the approximations used here.}
 \label{opconf}
\end{figure}

\begin{figure}
 \caption{Average order parameter
          $\langle \langle {\bar \varepsilon} \rangle \rangle$ vs. reduced
          temperature $t = (T - T_c^0) / T_c^0$ (a) (parameters appropriate for
          ${\rm Nb_3Sn}$);
          the region near $t_c(n_{\rm D}) = 0.444$ is magnified in (b).}
 \label{meanop}
\end{figure}

\begin{figure}
 \caption{Specific heat $C_v$ vs. reduced temperature $t$ (a) (parameters
          appropriate for ${\rm Nb_3Sn}$);
          the region near $t_c(n_{\rm D}) = 0.444$, where the discontinuity
          occurs, is magnified in (b).}
 \label{spheat}
\end{figure}

\begin{figure}
 \caption{Huang scattering intensity $S_c({\bf k})$ vs. reduced temperature $t$
          for different wave numbers $k = \zeta 2^{3/2} \pi / a_0$; the wave
          vector ${\bf k}$ lies in the soft sector.}
 \label{huasca}
\end{figure}

\end{document}